# VARIABILITATEA STRUCTURALA A PADURII NATURALE. STUDIU DE CAZ: CALIMANI

**Prof.univ.dr.ing. Radu CENUSA[1], Cercet.pr.I.dr.ing. Iovu BIRIS[2], Conf.univ.dr.ing. Florin CLINOVSCHI[3], Asist. ing. Ionut Barnoaiea[4], Şef lucr.ing. Ciprian PALAGHIANU[5], Cercet. Pr.III. ing.Marius TEODOSIU[6]**

[1,3,4,5]Universitatea Ştefan cel Mare Suceava
Str. Universităţii nr.13, Suceava, RO 72099
[1]E-mail: raducenusa@usv.ro, [3]E-mail: clinovsc@fim.usv.ro,
[4]E-mail: ibarnoaie@usv.ro,[5]E-mail: cpalaghianu@usv.ro

[2,6]Institutul de Cercetări şi Amenajări Silvice
Şos. Ştefăneşti nr. 128, Voluntari, Ilfov, RO 077190
[2]E-mail: ecologie@icas.ro, [6]E-mail: mcteodosiu@yahoo.com

## REZUMAT

*The paper presents the importance of research which characterizes the natural forest structure for the forest management. The lessons learned in these particular forest ecosystems can be integrated by the forest management objectives, in order to increase the sustainability of this type of resources. The project NATFORMAN was focused on the structure of the natural forest, thus research methodologies and modern technology (such as Field-Map) investigation and determination were used in order to record information on forest structural parameters. The results obtained refer to these structural parameters and to the possibility of transferring such information in practice, in order to achieve forest sustainable management.*

*Lucrarea prezintă importanţa cercetărilor de structură pentru caracterizarea pădurii naturale şi pentru transferul învăţămintelor desprinse din funcţionarea acestor ecosisteme forestiere către o gestiune silvică care se doreşte a fi din ce în ce mai mult legată de valorificarea durabilă a resurselor. În cadrul proiectului NATFORMAN, pe latura cercetării structurii a pădurii au fost utilizate metodologii moderne de investigare precum tehnologia Field-Map pentru determinarea şi stocarea informaţiilor privind parametri structurali ai pădurii. Rezultatele obţinute se referă la aceşti parametri structurali, la modul de transfer în practica silvică pentru realizarea gospodăririi durabile a pădurii.*

## 1 INTRODUCERE

În cadrul complex al ecosistemului forestier, arboretul reprezintă el însuşi un sistem termodinamic deschis, condus prin mecanisme biocibernetice de reglaj. Oricărui sistem cibernetic i se asociază noţiunea de structură, carcterizată ca o mulţime de entităţi esenţiale între care există o relaţie de ordine [1]. Cercetarea structurilor complexe presupune utilizarea unui instumentar sofisicat capabil să pună în evidenţă unele trăsături cât mai apropiate de un ideal, întrucât aproape întotdeauna se intervine cu delimitări şi cu descrieri menite a îndeplini mai mult sau mai puţin obiectivele cercetărilor subordonate unui anumit scop. În spatele acestora, rămâne **structura reală,** ale cărei numeroase atribute generează o imensă diversitate, laturi care se pot constitui în tot atât de multe şi diverse domenii în care cercetării ştiinţifice îi rămân porţi deschise, oricât de multe şi de mari ar fi progresele ştiinţei. Structura totală a ecosistemului este determinată de **forţe integratoare** care generează planul structural [2]. Din interacţiunea acestor forţe, rezultă **hipervolumul cu n dimensiuni al ecosistemului**, în spaţiul căruia se desfăşoară diferitele planuri structurale: structura de biotop, structura biocenotică, structura genetică, structura trofodinamică, structura biochimică, structura spaţială, etc. Aceste faţete ale structurii nu pot prezenta o expresivitate ridicată dacă nu sunt asociate cu **timpul,** de aici rezultând necesitatea cercetării dinamicii lor în timp şi spaţiu. Determinarea trăsăturilor structurale ale pădurilor prezintă o importanţă deosebită, întrucât aduce informaţii asupra gradului de naturalitate, al modului de manifestare şi al impactului factorilor ecologici perturbatori, precum şi a funcţionalităţii şi a echilibrului dinamic. Pădurile naturale prezintă un interes crescut din acest punct de vedere, întrucât structura lor se poate manifesta de cele mai multe ori ca rezultat al acţiunii forţelor integratoare ale naturii, fără intervenţia vizibilă a factorului uman. Proiectul NATFORMAN are ca principale obiective: implementarea reţelei de monitorizare a dinamicii structurale în ecosisteme forestiere reprezentative sub raportul conservării stării naturale; analiza principalilor parametri structurali şi a indicatorilor dinamicii structurii pădurii; caracterizarea cantitativă şi calitativă a lemnului mort şi a rolului acestuia în diferite tipuri de ecosisteme forestiere; evaluarea generală a biodiversităţii; formularea de recomandări pentru practica silvică rezultate din studiul pădurii naturale; diseminarea rezultatelor cercetărilor, în vederea gospodăririi durabile a pădurilor.

## 2 METODA DE LUCRU. LOCUL CERCETĂRILOR

O metodă adecvată pentru caracterizarea structurii spaţiale, care să permită o infuzie mai accelerată în teoria şi practica silvică a principiilor şi conceptelor care vizează cercetarea pădurilor naturale, este aceea ce se bazează pe teoria fazelor de dezvoltare. O fază de dezvoltare este o “etapă de dezvoltare evident diferenţiată structural în cadrul unei anumite asociaţii forestiere “ [3]. Astfel cercetarea structurii pădurii naturale, trebuie să aibă în vedere câteva cerinţe elementare:

- formele structurale tipice se stabilesc şi se caracterizează prin:amestecul de specii, etajarea pădurii, agregarea indivizilor, constituirea volumului lemnos, vitalitatea individuală, morfologia coroanei, tendinţele de evoluţie;
- analiza inelului anual trebuie să valorifice modelele auxometrice, pentru a evidenţia particularităţile elementelor structurale ;
- analiza dinamicii structurale trebuie să se întreprindă cu ajutorul instrumentelor caracteristice studiului fazelor de dezvoltare şi a succesiunii ecologice.

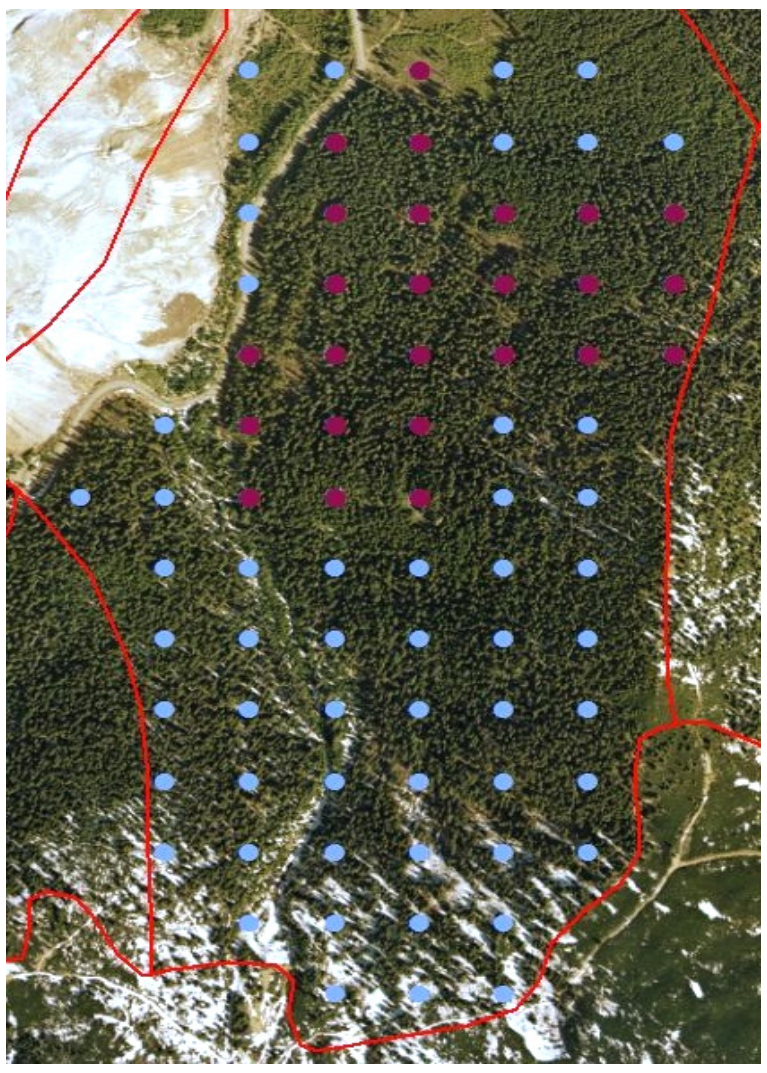

**Fig.1:** Reţeaua de puncte de inventariere pentru u.a. 69 A, Călimani

Pentru o bună reprezentare spaţială, inventarierile s-au efectuat într-o reţea de puncte amplasate geometric cu ajutorul GPS (fig.1), punctele fiind materializate pe teren cu ajutorul, unor piese metalice pentrua fi depistate mai uşor la reinventariere. Aria suprafeţei de probă a fost de 500 $m^2$ desimea reţelei depinzând de mărimea parcelei, (pentru Călimani au fost amplasate 75 de suprafeţe la o distanţă de 100 m.)

Pentru determinări asupra parametrilor structurali a fost utilizată tehnologia Field-Map, care permite determinarea caracteristicilor de bază ale suprafeţelor de probă (altitudine, expoziţie, pantă, coordonatele geografice ale centrelor) ale arborilor (poziţie, diametrul, înălţime, înălţimea elagată, starea de vegetaţie etc.); ale lemnului mort şi ale cioatelor. Fişa de ieşire cu cele 4 module ce permit determinarea fazelor de dezvoltare şi constituirea bazei de date, a monitoringului stării rezervaţiei forestiere (arbori pe picior, distribuţia spaţială pe specii, distribuţia spaţială după diametrul de bază, distribuţia spaţială după starea de vegetaţie şi poziţia lemnului mort) este prezentată în figura 2.

**Cercul 4** Altitudine: 1509 m Expoziţie: NV

Arbori pe picior

| Specie | Stare de vegetaţie | N (arbori/ha) | G($m^2$/ha) | V($m^3$/ha) |
|---|---|---|---|---|
| | | 820 | 48,937 | 312,796 |
| Mo | 1 | 1040 | 34,037 | 3,842 |
| | 2 | 160 | 0,842 | 121,115 |
| | 3 | 140 | 10,645 | 67,535 |
| | 4 | 80 | 8,495 | 505,288 |
| Mo Total | | 1420 | 54,020 | 505,288 |

Mediile şi coeficienţii de variaţie ai diametrului şi înălţimii

| Caracteristica | medie | s% |
|---|---|---|
| d | 18,2 | 69,01 |
| h | 13,4 | 55,56 |

| Volum lemn pe picior ($m^3$) | Volum lemn mort ($m^3$) | Volum total cerc ($m^3$) | Procent lemn mort (%) |
|---|---|---|---|
| 23,413 | 14,402 | 37,814 | 38,08 |

Distributia arborilor pe specii

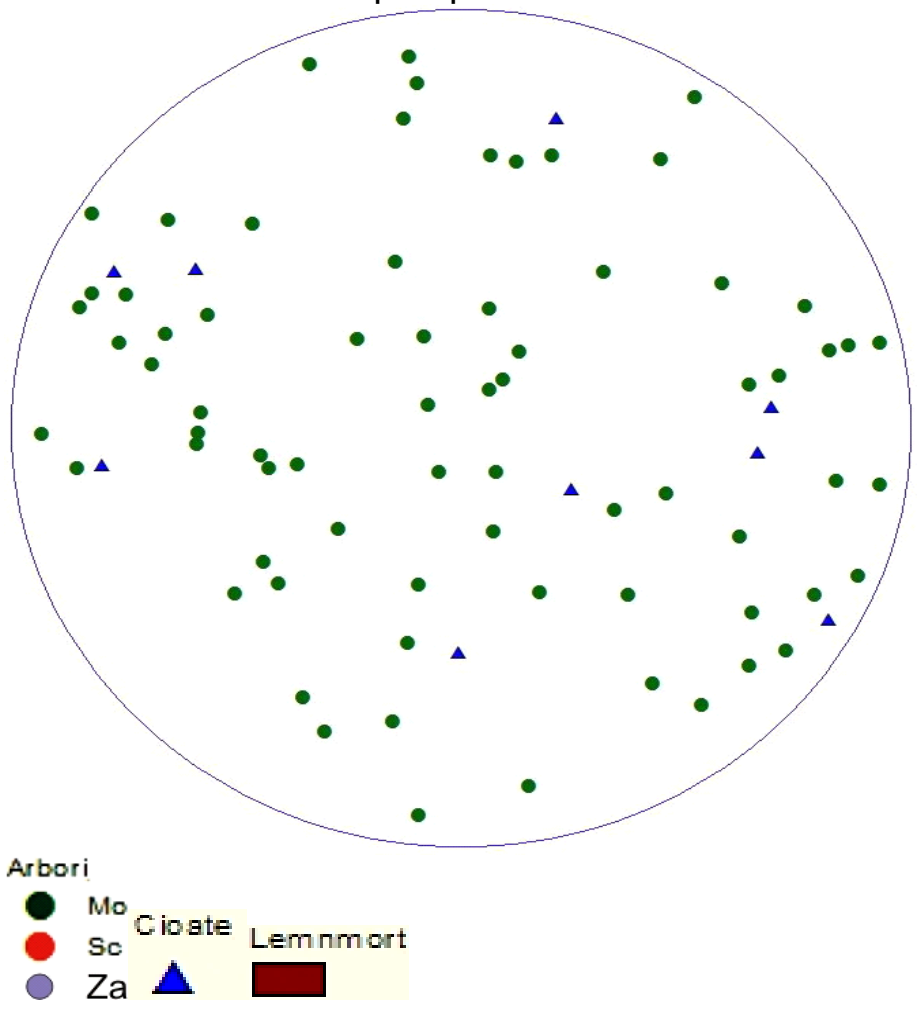

Variaţia diametrului de bază şi al diametrului cioatei în suprafaţă

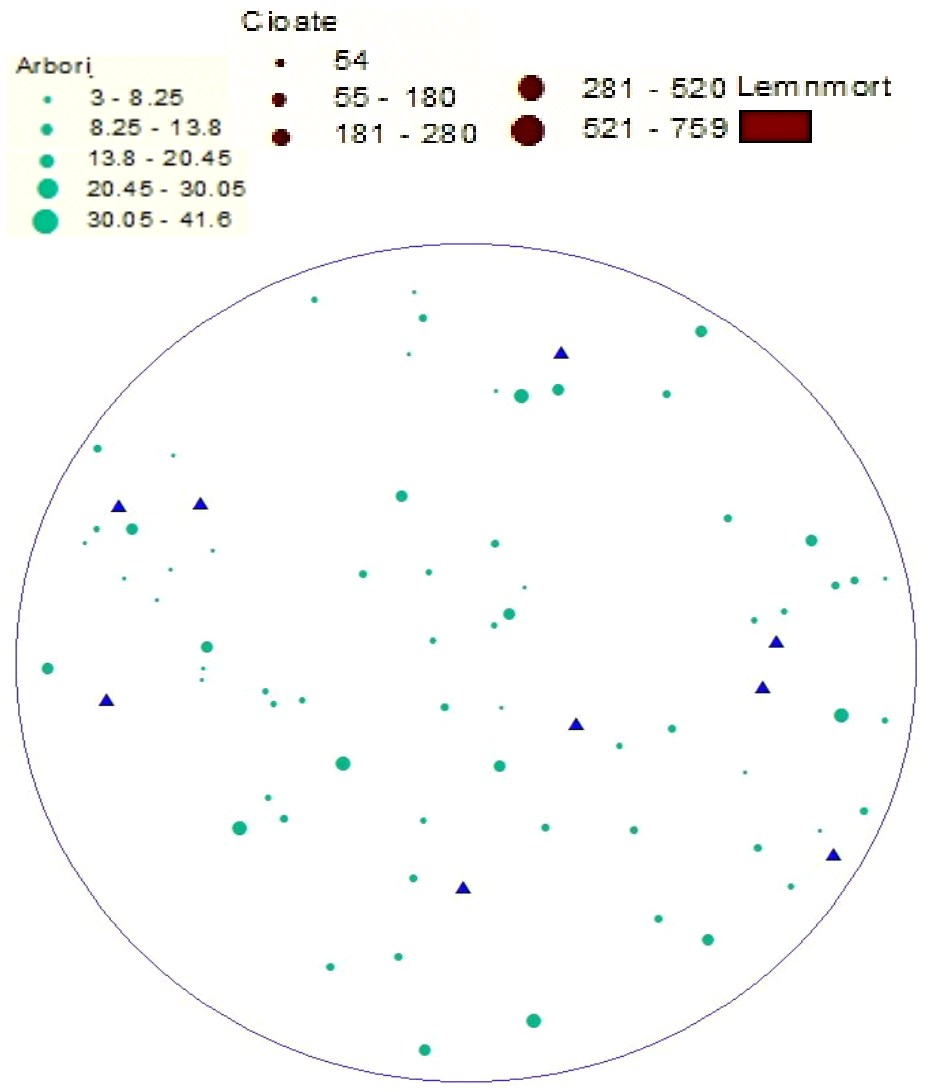

Distribuţia spaţială a arborilor după stare de vegetaţie şi a cioatelor şi lemnului mort după gradul de descompunere

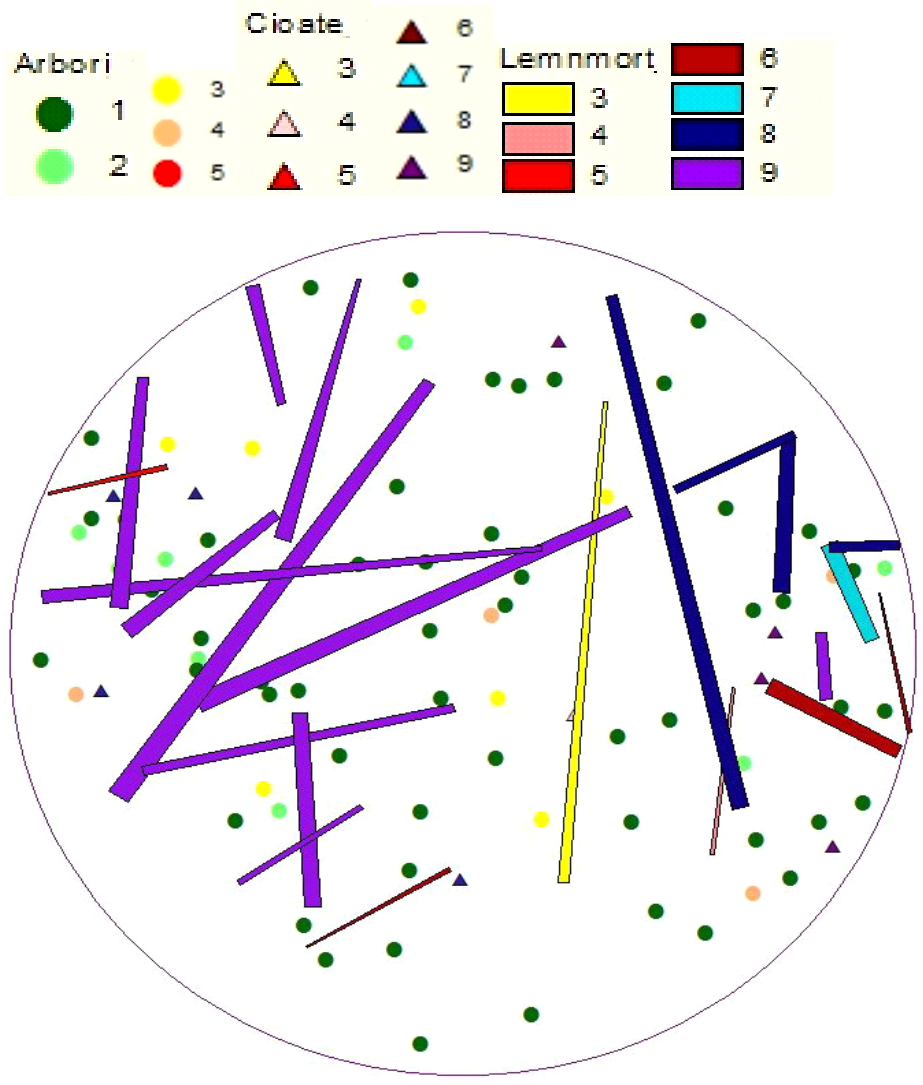

Faza de dezvoltare: Terminală cu regenerare

**FIG. 2**: Conţinutul fişei de inventariere şi de stocare a informaţiilor primare rezultat prin aplicarea procedeului de inventariere Field-Map

Cercetările din cadrul proiectului au fost efectuate în cinci rezervaţii forestiere naturale : Nera (Caraş-Severin), Runcu-Groşi (Arad), Slătioara, Giumalău şi Călimani (Suceava).

## 3 REZULTATE ŞI DISCUŢII

Prezentul capitol se referă, din raţiuni de spaţiu la cercetările desfăşurate în rezervaţia ştiinţifică *molidiş cu pin cembra* din cadrul parcului naţional Călimani.

Element esenţial al structurii orizontale şi verticale, desimea stratului arborescent poate să se constituie într-un parametru de bază în carcterizarea unei faze de dezvoltare, alături de diametru , înălţime şi de variaţia acestora. O imagine sintetică este prezentată în tabelul 1.

Tabelul 1. Principalele caracteristici structurale ale fazelor de dezvoltare

| Faza de dezvoltare | Suprafaţa ocupată –%- | Număr arbori/ha | Diametrul | | Înălţimea | |
|---|---|---|---|---|---|---|
| | | | Mediu -cm- | Coef.var. -%- | Medie -m- | Coef.var. -%- |
| Iniţială | 12 | 1546 | 18,6 | 58,6 | 15,6 | 47,6 |
| Optimală | 16 | 1000 | 27,1 | 41,4 | 19 | 40,4 |
| Optimală târzie | 16 | 665 | 27,5 | 47,9 | 19,3 | 41,0 |
| Terminală cu regenerare | 24 | 913 | 22,5 | 65,1 | 14,5 | 52,1 |
| Degradare | 8 | 320 | 30,5 | 54,5 | 20,3 | 45,6 |
| Degradare cu regenerare | 12 | 567 | 29,5 | 56,2 | 16,1 | 62,4 |
| Regenerare | 12 | 407 | 16,4 | 49,9 | 8,9 | 38,3 |

Pentru suprafaţa inventariată, numărul mediu de arbori este de 782, fiind răspândit neuniform. Amplitudinea acestei neuniformităşi este dată de un coeficient de variaţie de 51,7%. Specia cu constanţa spaţială cea mai ridicată este molidul, cu 716 arbori/ha şi un coeficient de variaţie de 50% spre deosebire de zâmbrul care are o reprezentare mai modestă (46 exemplare/ha) şi o raspândire foarte neuniformă (coeficient de variaţie de 205%) şi de scoruş, specie al cărei caracter pionier se manifestă printr-o prezenţă sporadică (18 exemplare/ha şi un coeficient de variaţie de 304%). Neuniformitatea răspândirii este ilustrată şi din datele din tabel din care rezultă că în fazele tinere numărul de arbori este cel mai ridicat. Variabilitatea structurală se regăseşte şi la nivel dimensional, diametrul şi înălţimea medie fiind caracteristice fiecărei faze, la fel ca şi coeficienţii de variaţie.

Analiza structurii fazelor de dezvoltare pune în evidenţă alternanţa dintre procesele de uniformizare şi cele de diversificare structurală, precum şi unitatea dintre brusc şi lent, dintre continuu şi discret, dintre regenerare şi eliminarea naturală. Aceste aspecte ilustrează caracterul complex, holistic al relaţiilor care există între subsistem (arbore) şi sistem (pădurea naturală), între individ şi colectivitate. Evoluţia ciclică a fazelor de dezvoltare are traiectorii proprii, în raport cu caracteristicile staţionale şi cutipul de ecosistem. Alternanţa dintre structurile simple şi cele complexe este parte integrantă a acestei evoluţii, pe care se bazează realizarea echilbrului dinamic. După cum se cunoaşte, pentru structura pădurii naturale, studiul cantitativ şi calitativ al necromasei lemnoase, reprezintă o etapă de importanţă deosebită. Sub raportul volumului, pentru suprafaţa Calimani, rezultatele sunt prezentate în tabelul 2.

Tabelul 2. Structura volumului în raport cu fazele de dezvoltare

| Faza de dezvoltare | Volum lemn -m³- | | | % lemn mort |
|---|---|---|---|---|
| | Total | Pe picior | Lemn mort | din volumul total |
| Iniţială | 812 | 742 | 71 | 8,7 |
| Optimală | 570 | 403 | 167 | 29,2 |
| Optimală târzie | 517 | 442 | 75 | 14,6 |
| Terminală cu regenerare | 605 | 486 | 119 | 19,6 |
| Degradare | 336 | 228 | 108 | 32,1 |
| Degradare cu regenerare | 528 | 342 | 186 | 35,2 |
| Regenerare | 146 | 128 | 17 | 12,1 |

Se constată mari fluctuaţii ale volumului arbirilor pe picior (amplitudine de 612m$^3$ între faza iniţială şi faza de regenerare). Totodată, se poate vedea că volumul necromasei poate deţine în funcţie de faza de dezvoltare, până la 35% din volumul total. De altfel s-a identificat o relaţie directă pozitivă între volumul lemnului mort şi diametrul mediu (r = 0,406*) care arată că o dată cu acumularea biomasei se produce şi acumularea de lemn mort. O altă legătură de acelaşi tip s-a constatat între volumul lemnului mort şi numărul de puieţi la unitatea de suprafaţă ( r = 0,362*). Această relaţie conduce la ipoteza că lemnul mort reprezintă şi un suport pentru germinaţia seminţelor şi menţinerea unor condiţii favorabile pentru puieţi.

Seminţişul, un alt element care joacă un rol de bioindicator al gradului de naturalitate este reprezentat în principal de molid şi de scoruş, zâmbrul este semnalat cu totul sporadic. Fapt normal pentru o pădure închisă, cantitatea de seminţiş nu atinge decât 0.68 puieţi/m$^2$, având o răspândire neuniformă (coeficientul de variaţie este de 96%). Molidul este răspândit mult mai uniform (CV: 66%) decât scoruşul care este mai neuniform (CV:160%), dar mai abundent.

Covorul erbaceu este foarte bine reprezentat de afin (*Vaccinium myrtillus*). Speciile comune ale acestui strat care arată o răspândire constantă şi relativ uniformă sunt: *Oxalis acetosella, Homogyne alpina, Luzula sylvatica, Deschampsia flexuosa, Calamagrostis villosa, Poa nemoralis, Festuca supina*, precum şi ferigile *Dryopteris dilatata, Athyrium filix-femina*. Lor li se mai adaugă şi alte specii cu o frecvenţă ceva mai redusă: *Luluza luzuloides, Soldanella hungarica, Festuca rubra, Moneses uniflora, Campanula abietina, Rumex arifolius, Senecio nemorensis* ssp. *fuchsii, Deschampsia caespitosa, Dryopteris filix-mas, Dryopteris carthusiana, Huperzia selago, Athyrium distentifolium, Dryopteris disjuncta, Lycopodium annotinum.*

În cuprinsul stratului erbaceu sau imediat sub el, apar frecvent speciile de muşchi, între care se remarcă: *Hylocomium splendens, Pleurozium schreberi, Dicranium scoparium,* iar mai rar apar şi *Dicranium fuscescens, Rhytidiadelphus triqueter, Polytrichum juniperinum, Polytrichum formosum*.

# 4 CONCLUZII

Existenţa simultană pe spaţii restrânse a unui număr suficient de mare de faze de dezvoltare, cu o funcţie specifică, îi conferă biocenozei un grad ridicat de integralitate şi în consecinţă o capacitate ridicată de reacţie la fluctuaţiile externe pertubatoare. Desigur există limitele amplitudinale ale fluctuaţiilor, care o dată depăşite lasă loc ireversibilităţii şi, trecerii sistemului pe alte traiectorii. În consecinţă, întreaga evoluţie a pădurii naturale, depinde de momentul şi de intensitatea acţunii perturbării, precum şi de complexitatea mozaicului structural funcţional.

Pădurea naturală dispune de mecanisme structurale bine definite menite să asigure un echilibru dinamic ca principal rezultat al adaptării viului la condiţiile staţionale şi la regimul factorilor perturbatori. Strategiile pădurii naturale nu sunt concordante cu unele din obiectivele pe care sistemul social-global le fixează gospodăriei silvice prin ţelurile de gospodărire.

Pădurea cultivată, cu structură echienă, monoetajată şi de cele mai multe ori monospecifică, poate să manifeste la un moment dat, una sau cel mult două funcţii ecosistemice (de acumulare de biomasă, de regenerare, de restructurare) în timp ce pădurea naturală, dispunând de mozaicul structural complex, exercită concomitent, pe un spaţiu restrâns toate funcţiile ecosistemului forestier.

Pădurea naturală, face o mare „risipă" de biomasă, angrenată în circuitele biogeochimice care în pădurea cultivată nu s-ar justifica din punct de vedere economic. Dar această biomasă constituie un element dinamizator al pădurii, reprezentând nuclee de regenerare pe de-o parte, şi microhabitate pentru diverse organisme vegetale şi animale, de cealaltă parte.

Manifestarea, violentă uneori, a factorilor perturbatori, este caracterizată în cazul pădurilor cultivate, drept catastrofă. Pentru pădurea naturală însă, ea face parte dintr-un ciclu normal, menit să asigure echilibrul dinamic care are drept scop autoperpetuarea pădurii.

# BIBLIOGRAFIE